\documentclass{article}    
\usepackage{graphicx}
\usepackage{amssymb}

\begin{document}           

\title{Niederhauser's Model for Epilepsy \\
and Wavelet Methods}  
\author{J. P. Trevi\~no$^1$, 
V. H. Castillo$^1$, H.C. Rosu$^1$, J. L. Mor\'an L\'opez$^1$, 
\\ and J. S. Murgu\'{\i}a$^2$ \\ 
{$^1$\it  \small IPICyT - Instituto Potosino de Investigaci\'on Cient\'{\i}fica y Tecnol\'ogica,}\\
{\it \small Apdo Postal 3-74 Tangamanga, 78231 San Luis Potos\'{\i}, S.L.P., M\'exico.}\\
{$^2$ \it \small 
Universidad Aut\'onoma de San Luis Potos\'{\i}, 87545 San Luis
Potos\'{\i}, S.L.P., M\'exico.}\\
 {\it \small emails:  jpablo, vcastillo, hcr,
moran-lopez@ipicyt.edu.mx, ondeleto@uaslp.mx }}

\maketitle                 

Wavelets and wavelet transforms (WT) could be a very useful tool to
analyze electroencephalogram (EEG) signals. To illustrate the WT
method we make use of a simple electric circuit model introduced by
Niederhauser \cite{semey}, which is used to produce EEG-like
signals, particularly during an epileptic seizure. The original
model is modified to resemble the 10-20 derivation of the EEG
measurements. WT is used to study the main features of these
signals.

\section{Brain and Neurons}
The body of animals, including the human being, is controlled by the
nervous system. This system has a primary division: central and
peripheral. The brain, or cerebrum, the cerebelum, and the spinal
cord form the central nervous system, while the peripheral structure
is integrated by long nerves that reach every part of the body. The
brain is organized in zones which perform specific tasks, which
nowadays are the subject of more detailed studies.

At the microscopic scale, the basic functional units of the brain
and nerves are a class of excitable cells called neurons. The human
brain alone contains about $10^{11}$ neurons which come in different
shapes and sizes but have the same general morphology. The soma
(body) of a neuron can measure from 1 $\mu$m up to 1 mm across,
contains the nucleus, and has two main sets of membrane elongations:

(i) dendrites are prolongations through which the neuron receives
information from other neurons

(ii) the axon is the main prolongation through which a neuron sends
signals to the outside.

\begin{figure}[htbp]
   \centering
   \includegraphics[width=4.8cm]{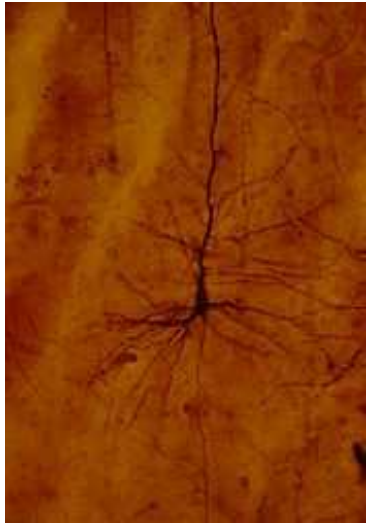}
   \caption{This is a photomicrograph from Cajal's preparations (housed in the Museo Cajal at the Cajal Institute, Madrid, Spain) of a
   neuron from the cerebral cortex of a newborn infant, impregnated by the Golgi stain. The soma, the axon, and the dendrites are easy to identify.
   Courtesy of http://nobelprize.org/medicine/articles/cajal/}
\end{figure}

\section{Brain Cortex}
A well-defined spatial organization of the human brain is through
stacks of layers. The outermost layer is the cortex in which many of
the higher activities are performed: memory, attention, perceptual
consciousness, thought, and language. This layer is about 3 mm only,
but despite the small dimension is of basic interest in the research
of neuro-physiologists because it engenders important features of
the human thought. The cortex is usually studied through the
electric signals that produces. The way to detect these signals is
through the electroencephalogram (EEG), which is basically a record
of the electric activity as obtained by electrodes on the scalp. The
EEG could be understood as a superposition of the individual signals
coming from each neuron in a given lapse of time \cite{liley}. This
makes it a very useful tool as an experimental surface measure of
the activity of a certain number of neurons that are of specific
interest \cite{science04}.

\subsection{Basics of EEG}
The history of EEG begins in 1875 when Richard Caton (1842-1926) in
Liverpool discovered the existence of electrical signals from the
exposed  brain of rabbits and monkeys. This discovery was done by
employing the galvanometer invented seventeen years earlier by Lord
Kelvin. Later, in 1913, the Russian physiologist V. V.
Pradvich-Neminsky published the first EEG ever recorded from a dog.
At the present time, the EEG is one of the most important methods
for the study of neural activity at the level of the brain cortex.
It is usually a tool for diagnosis of several important disorders
such as autism, language problems, and epilepsy, as well as motor
damages.

To obtain EEG data, electrodes should be positioned onto the scalp
of the patient. The distribution of the electrodes along with the
reference used to measure the signal is called derivation. Though
there are several types of derivations, the most commonly used is
the 10-20 one. Its name comes form the fact that the electrode
arrangement is referred to proportions of skull measures ($10\%,
20\%$ and so on).

\begin{figure}[htbp]
   \centering
   \includegraphics[width=6cm]{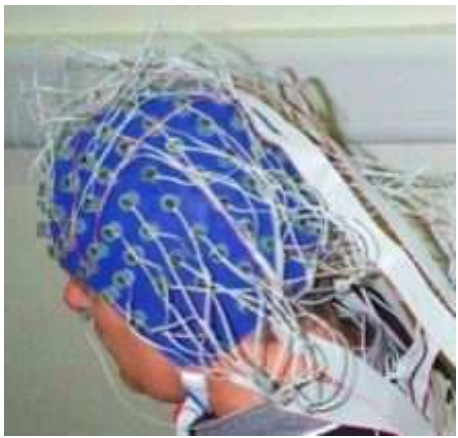}
   \caption{The 10-20 configuration is the most common to obtain EEG data for diagnosis. The number of electrodes depends on the equipment
   available and the required precision.}
   \label{fig:eeg}
\end{figure}

\section{Niederhauser's Model of Epilepsy}

In the original setup of the EEG, the signal is sent from the scalp
to moving needles which record it on a sheet of paper. Nowadays,
experts make use of samplers and computers to create data sets to
represent the EEG as a set of channels which resemble the usual EEG.
Since the EEG is a set of time series which reflect the activity of
different groups of neurons, it is possible to describe the behavior
of a cluster of neurons with a few simple interaction rules. This
was the basic idea of Niederhauser \cite{semey} who proposed a
discrete model on which we will focus in the following. The model
takes into account basic features of real neurons to produce an EEG
like signal at normal periods and also through the so-called
epileptic seizures, which roughly means a sudden start of a regime
of strong oscillations.

On the other hand, epilepsy is a very complicated disease and has
different manifestations. The model proposed by Niederhauser is thus
referred solely to epileptic seizures possessing apparent dominant
frequencies. These crises are associated to the hypersynchrony of
large groups of neurons and some degree of order is considered
theoretically. The neuronal units (called neuronions) are
distributed within a rectangular zone array with a set of simple
interaction rules. The neuronions are programmed to transport
electric charge from one zone to another in a conditional way when a
threshold charge difference is reached. If the charge difference is
below the threshold, the neuronion has only a small probability to
fire. The target zone of each neuron is random and a threshold value
for the charge transportation has to be set up at the beginning of
the simulation.

\begin{figure}[htbp]
   \centering
   \includegraphics[width=9cm]{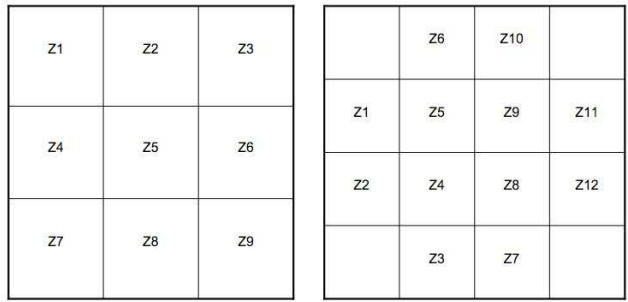}
   \caption{ Left: the original Niederhauser's configuration. Right: the 10-20 configuration.}
   \label{fig:zonas}
\end{figure}

\begin{figure}[htbp]
   \centering
   \includegraphics[width=10.4cm]{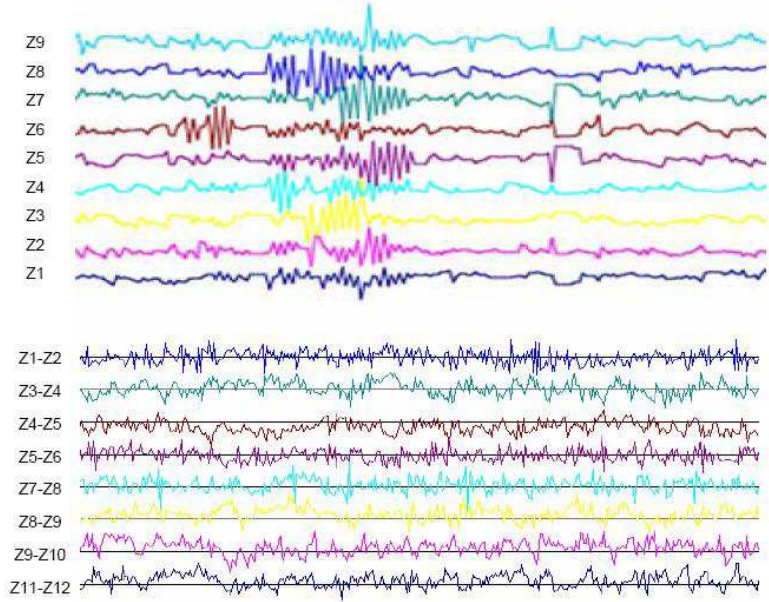}
   \caption{Output signals generated by our simulations of the original Niederhauser's model (top) and
   the 10-20 derivation (bottom).}
   \label{fig:signals}
\end{figure}

The original model considers $2\cdot 10^4$ neuronions distributed
over nine regions in a rectangular 3 $\times$ 3 array as shown in
fig. \ref{fig:zonas}. The most important parameter of the model is
the threshold voltage which is in direct correspondence with the
firing threshold of a real neuron. If a large threshold value is
chosen, the output of the simulation will resemble a normal EEG
signal, whereas a small threshold value will yield a seizure-like
output. To make the original model more realistic, a larger number
of neurons were distributed over a 4 $\times$ 4 arrangement shown in
fig. \ref{fig:zonas} that fits better the simplest 10-20 derivation.
We found that the output signal did not change significantly (see
fig.~\ref{fig:signals}), which means that the configuration of the
zones is not a critical parameter in the modeling.

\section{Wavelet Theory}
Wavelet transforms (WT) are generalized Fourier transforms that in
the last two decades have been extensively used to investigate
special features of real functions such as scalar one dimensional
fields, and more usually, time series. The WT has significant
advantages over the common Fourier transforms. The most simple way
to argue in favor of the WT is that, unlike the non-localized
Fourier spectrum, the WT gives details of the signal at different
resolutions and portions of the entire signal.

In general, wavelets are functions in the class $\psi(t) \in
L^2(\mathbb{R})$ with the following properties:
\begin{eqnarray}
C_\psi&=& \int_0^\infty \frac{|\hat \psi(\omega)|^2}{|\omega|}d\omega<\infty\\
\int_{-\infty}^\infty\psi(t)dt&=&\hat \psi(0)=0,
\end{eqnarray}
where $\hat \psi(\omega) = \int e^{i\omega t} \psi(t) dt$ is the
Fourier transform of $\psi(t)$.

The first equation is an admissibility condition, while the second
one is the zero mean condition. The function $\psi(t)$, known as the
mother wavelet, can be used to build an orthonormal basis of
translated and dilated functions of the form

\begin{equation}
\psi_{a,b}(t)=\frac{1}{\sqrt{a}}\psi\left(\frac{t-b}{a}\right)~.
\end{equation}

The WT of a function $f$ that we denote by $\hat f_{a,b}(t)$ is
defined as the scalar product in $L^2(\mathbb{R})$ of the function
with the chosen wavelet:
\begin{equation}
\hat f_{a,b}(t)=\langle f,\psi_{a,b} \rangle.
\end{equation}
The WT measures the variation of $f$ in a neighborhood of size
proportional to $a$ centered on point $b$.

One fundamental property that is required in order to analyze
singular behavior is that $\psi(t)$ has enough
vanishing moments. 
A wavelet is said to have $n$ vanishing moments
  if and only if it satisfies

\begin{eqnarray}
  \int_{-\infty}^{\infty} t^k \psi(t) dx\ &=&\ 0,\ {{\rm for}\ k = 0, 1, \ldots , n - 1}
\end{eqnarray}
and
\begin{eqnarray}
  \int_{-\infty}^{\infty} t^k \psi(t) dt\ &\neq&\ 0, \ {{\rm for}\ k \geq n.} \label{momentos}
\end{eqnarray}

  This means that a wavelet with $n$ vanishing moments is orthogonal to
  polynomials up to order $n-1$. In fact, the admissibility condition requires at least one vanishing moment. So the wavelet
  transform of $f(t)$ with a wavelet $\psi(t)$ with $n$ vanishing moments
  is nothing but a ``smoothed version'' of the $n$th derivative of
  $f(t)$ on various scales. When one is interested to
  measure the local regularity of a signal this
  concept is crucial. In the plots of fig.~5 we used Daubechies wavelets with 8 and 20 vanishing moments, respectively.

As the set of wavelets form a basis, any function can be decomposed into the linear combination

\begin{equation}
f(t)=\sum_m\sum_n x_n^m\psi_{m,n}(t),
\end{equation}
with coefficients
\begin{equation}
x_n^m=\int_{-\infty}^\infty f(t)\psi_{m,n}(t)dt,
\end{equation}
where the basis functions are defined in terms of the mother wavelet
as follows
\begin{equation}
\psi_{m,n}(t)=2^{m/2}\psi(2^mt-n).
\end{equation}

In the wavelet approach the fractal character of a certain signal
can be inferred from the behavior of its power spectrum $P(\omega)$,
which is the Fourier transform of the autocovariance (also termed
autocorrelation) function and in differential form
$P(\omega)d\omega$ represents the contribution to the variance of a
signal from frequencies between $\omega$ and $\omega+d\omega$.

Indeed, it is known that for self-similar random processes the
spectral behavior of the power spectrum is given by \cite{wor,stas}

\begin{equation}\label{wvl_4}
P_{\varphi}(\omega)\sim |\omega|^{-\gamma _f}~,
\end{equation}
where $\gamma _f$ is the spectral parameter of the wave signal. 

\begin{figure}[htbp]
   \centering
   \includegraphics[width=11cm]{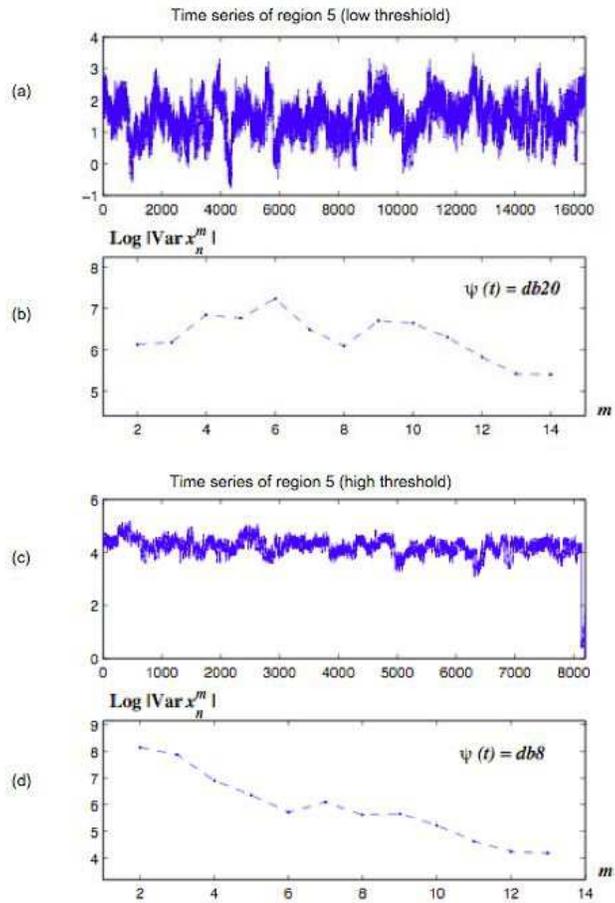}
   \caption{The time series above correspond to the 5th zone of an original Niederhauser's arrangement. Plots (a) and
   (b):
   an epileptic seizure event simulation is shown for which the wavelet coefficients do not display a self-similar fractal structure.
    Plots (c) and (d):
   a normal EEG simulation is shown for which the wavelet coefficients could be argued to have a fractal behaviour.
}
   \label{fig:ondeleta}
\end{figure}

In addition, the variance of the wavelet coefficients possesses the
following behavior \cite{stas}
\begin{equation}
{\rm var} \,x _{n}^{m} \approx \left(2^{{ m}}\right)^{-\gamma _f}~.
\end{equation}

These results will be employed to study the output of the
Niederhauser model, and also of real EEG data for comparison
purposes.

\section{WT and EEG Signals}

It is relatively easy to use wavelet theory to analyze EEG data,
although the interpretation of the results is not so easy.  There is
previous work that links the time series analysis through the WT to
the analysis of EEG data. The detection of  the so called epileptic
spikes is explained in \cite{latka}, where the authors also mention
a comparison between this method and the available software within
the medical community. In this work, wavelet theory will be applied
to the model by Niederhauser in the particular case of epilepsy.

The normal EEG is sometimes thought of as a chaotic signal. There is
some discussion about this issue in \cite{inability} and previous
works, where  Lyapunov exponents theory is used to measure chaos.
Wavelet analysis provides a simple algorithm to determine the
fractal dimension (closely related to the Hausdorff-Besicovich
dimension) of a curve, and therefore conclude whether it is a
fractal or not.

In fig.~5 simulations of the EEG with the original Niederhauser's
model are shown. The wavelet coefficients reveal a fractal behaviour
in the normal EEG while in the epileptic seizure, the coefficients
cannot give us such information. We find that during the epileptic
seizure there is a dominance of a given scale, which could be
interpreted as an ordering of neurons at a certain scale.

The behaviour of the output of the modified system is qualitatively
the same. This means that the fractal and nonfractal feature of the
respective episodes are constant. From these computations one could
conclude that the normal EEG has a fractal feature. Despite these
results, the same analysis for real EEG is missing, though the same
results are expected. Additionally, the results of the simulations
support the idea that at an epileptic seizure there is some degree
of order in the EEG signals.

\section{ Conclusions}
We reconsidered the simple electric circuit model of Niederhauser
for epilepsy with minor modifications. We confirm that it is capable
to reproduce specific features of EEG data such as frequency or
scale dominance at a seizure and fractality at normal periods. This
model is useful for checking different methods for EEG signal
analysis and gives insight to non-medic students on certain basic
features of epilepsy. It could even give a clue of the causes and
behaviour of the disease itself if appropriate modifications are
performed. As an example, we used wavelet transform analysis since
we believe it could be a useful tool in getting a wealth of
information about particular features of the EEG signals from
pathological conditions in different patients to specific details
about a given patient. In the future, we hope to make further
modifications of the model and the analysis of the data through
wavelet analysis to seek for more details of epileptic disorders and
their relationships to neuronal dynamical features at the level of
the whole brain. A software development with characteristics similar
to current software, such as spike detection, through wavelet
transform is under consideration.

\section{Acknowledgment}

This work was partially sponsored by grants from the Mexican Agency
{\em Consejo Nacional de Ciencia y Tecnolog\'{\i}a} through project
No. 46980-R.


\end{document}